\documentclass[conference,a4paper]{IEEEtran}
\IEEEoverridecommandlockouts
\usepackage{cite}
\usepackage{amsmath,amssymb,amsfonts}
\usepackage{algorithmic}
\usepackage{graphicx}
\usepackage{textcomp}
\usepackage{xcolor}
\usepackage{enumitem}
\usepackage{pgfplots}
\usepackage[hidelinks]{hyperref}
\usepackage{mathtools}
\usepackage{wrapfig}
\def\BibTeX{{\rm B\kern-.05em{\sc i\kern-.025em b}\kern-.08em
    T\kern-.1667em\lower.7ex\hbox{E}\kern-.125emX}}

\begin{document}

\title{Providing Curative Distribution Grid Flexibility Using Online Feedback Optimization}

\author{
    \IEEEauthorblockN{
        Florian Klein-Helmkamp\IEEEauthorrefmark{1},
        Fabian Böhm\IEEEauthorrefmark{1},
        Lukas Ortmann\IEEEauthorrefmark{2},
        Alexander Winkens\IEEEauthorrefmark{1},\\
        Florian Schmidtke\IEEEauthorrefmark{1},
        Saverio Bolognani\IEEEauthorrefmark{2},
        Florian Dörfler\IEEEauthorrefmark{2}
        and Andreas Ulbig\IEEEauthorrefmark{1}
    }
    \IEEEauthorblockA{
        \IEEEauthorrefmark{1}IAEW at RWTH Aachen University, Aachen, Germany
        \\\{f.klein-helmkamp,a.winkens,f.schmidtke,a.ulbig\}@iaew.rwth-aachen.de
        \\ fabian.jonathan.boehm@rwth-aachen.de
    }
    \IEEEauthorblockA{
        \IEEEauthorrefmark
        {2}Automatic Control Laboratory at ETH Zurich, Zurich, Switzerland
        \\\{ortmannl,bsaverio,dorfler\}@ethz.ch
    }}

\IEEEoverridecommandlockouts
\IEEEpubid{\makebox[\columnwidth]{979-8-3503-9678-2 /23/\$31.00~\copyright2023 IEEE \hfill} \hspace{\columnsep}\makebox[\columnwidth]{ }}

\maketitle

\IEEEpubidadjcol

\begin{abstract}
Distribution grid flexibility is regarded as a possible measure in curative system operation, yielding a need for an efficient and robust coordination mechanism for the joint flexibility provision by individual units to the transmission grid. This paper introduces a method to coordinate distribution grid level flexibility as a fast-responding curative measure based on Online Feedback Optimization. We utilize an optimization algorithm in a closed loop with the distribution grid to dispatch set points for active and reactive power to flexibility providing units. The approach is evaluated in an experimental setup, utilizing assets connected to an exemplary low voltage grid. Online Feedback Optimization is found to be both a viable as well as a highly effective approach to coordinate distributed energy resources in real-time curative system operation.
\end{abstract}


\section{Introduction}
\begin{figure}[t]
    \includegraphics[width=0.485\textwidth]{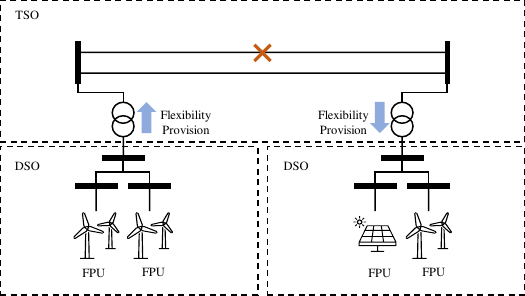}
    \caption{Provision of flexibility by distribution grids after an overload on a superimposed transmission line.}
    \label{fig:overview}
\end{figure}

Curative operation of energy transmission systems is a promising approach to allow for more efficient utilization of existing operating resources, therefore reducing the need for expensive preventive re-dispatch, while maintaining a high level of security in online system operation \cite{InnoSys}. Such curative operation becomes possible through fast-responding measures, that are able to reduce line loading below the permanently admissible limits (PATL) in case of an overload within a short interval of time, ranging from several seconds to several minutes. One curative measure under discussion is a provision of flexibility by distribution grids at the point of common coupling (PCC) between transmission system operator (TSO) and distribution system operator (DSO) as shown in \autoref{fig:overview}. In case of an overloaded transmission line, the requested flexibility has to be coordinated by the DSO to a possibly high number of flexibility providing units (FPU), which leads to a need for efficient and reliable coordination mechanisms.

The dispatch of set points for FPU can be formulated as an optimization problem preventing the violation of operational limits through constraints. Subsequently, it can be solved in an Optimal Power Flow (OPF) calculation. While OPF is a well-established approach for the generic economic dispatch problem, it also suffers from two major drawbacks that can be prohibitive for its use in curative system operation. First, complete and accurate models of distribution grids are often not available but are a necessity for OPF calculation. Second, a solution for the OPF is not guaranteed to satisfy all constraints of the power system if a mismatch between the model and the physical system exists. A promising approach to address these challenges is Online Feedback Optimization (OFO), i.e. implementing an optimization algorithm that acts as a controller in a closed loop with the power grid. To ensure the provision of flexibility in a use case that poses high requirements on availability and reliability, a novel approach like OFO has to be evaluated with regard to these requirements.

\subsection{Related Work}
An approach for aggregation of flexibility on the distribution grid level, considering time-dependencies in possible flexibility provision, is described in \cite{Contreras_2019}. The presented method is applied to the use case of flexibility provision at the TSO/DSO interface, considering requirements for ancillary services such as congestion management in superimposed grid layers. The analysis is performed under the assumption, that set points within the stationary limits, i.e. the feasible operating region (FOR) of the aggregated FPU, can be fully reached by the next time step considered in the simulation. Possible time limitations in flexibility provision due to ramp rates are described in \cite{Riaz_2019}. The FOR concept is expanded upon by the flexibility operating region (FXOR), i.e. the set of operating points at the PCC between TSO and DSO that are fully reachable within a certain time frame. Besides aggregation at the TSO/DSO interface utilizing flexibility necessitates coordination of requested active and reactive power to the FPU dispersed in underlying grids. A possible method based on OPF calculation is presented in \cite{Frueh_2023}. The specific use-case of distribution grid flexibility in curative system operation is introduced in \cite{Kolster_2020}. The presented results include key performance indicators (KPI), e.g. the availability of flexibility for curative congestion management, highlighting the increased requirements on reliability and safety for the given use-case. This is expanded upon in \cite{Kolster_2022}, where specific requirements for curative flexibility are formulated and a coordination approach based on a techno-economic dispatch is introduced. 

A common characteristic of the approaches to flexibility coordination is the strong reliance on accurate distribution grid models, that are being used in OPF and techno-economic dispatch calculations. Tackling this issue, controlling a dynamic system by utilizing optimization algorithms in a closed loop with online measurements from the system itself is a promising approach that is discussed in the literature for several different use cases \cite{Hau_2021}. A possible feedback optimization control scheme is described in \cite{Härberle_2020}. An approach to solve the AC OPF with OFO is introduced in \cite{Picallo_2023}. Applying the concept to voltage control in an exemplary power system, \cite{Ortmann_2020} describes a method to utilize OFO in ensuring operational limits for bus voltages. The applicability in grid operation was shown by implementing a central voltage controller in a laboratory environment. Further applicability for reactive power optimization has been underlined in \cite{Ortmann_2023} by deploying an OFO controller in an operational environment of a DSO.

\subsection{Main Contribution}
\label{subsec:RQ}
\begin{figure}[t]
    \includegraphics[width=0.485\textwidth]{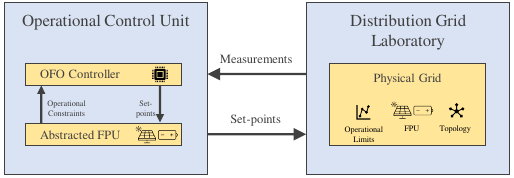}
    \caption{Interface between OFO-based operational control unit and physical distribution grid with FPU in laboratory setup.}
    \label{fig:method}
\end{figure}

This paper presents a method for coordinating and providing active power as flexibility at the PCC between two grid layers using an OFO controller. We describe a controller, optimizing the individual feed-in of FPU in a distribution grid while ensuring voltage limits. A request for flexibility is applied to the setup with the proposed controller adjusting the measured power flow at the PCC for the given set point. The applicability of OFO for the coordination of curative distribution grid flexibility is examined by implementing the algorithm in an experimental laboratory setup consisting of an exemplary low voltage distribution grid. We formulate the following research questions:
\begin{enumerate}[label=\roman*.]
    \item How fast can flexibility be provided by OFO?
    \item How are disturbances influencing the provision of flexibility coordinated by OFO?
\end{enumerate}
Based on previous stimulative studies, we expect OFO to be a viable approach with the achievable performance mainly being bounded due to hardware limitations, e.g. delay and sampling rate. To deploy the flexibility coordination with an OFO controller in the laboratory it is implemented as the central operating logic as shown in \autoref{fig:method}. The assets of the physical distribution grid are abstracted in an operational control unit (OCU) that serves as an interface to the physical hardware. It receives a set point for the active power flow at the PCC to the superimposed grid layer. OFO is subsequently optimizing the individual set points for active and reactive power of the FPU to provide the requested flexibility, while satisfying the operational constraints, i.e. voltage limits and power capabilities of FPU. Active and reactive power set points are processed and sent by the OCU to the physical FPU.

\section{Flexibility Provision with Online Feedback Optimization}
\label{sec:Modelling}
In the following section requirements for coordinating curative flexibility are presented to create a basis for the evaluation of the OFO algorithm. Subsequently, the chosen approach to solving the dispatch problem using OFO as a central controller is presented in detail.
 \subsection{Requirements for Curative Flexibility}
 \label{subsec:Requirements}
The evaluation of OFO as an approach to flexibility coordination for curative measures is performed with respect to typical requirements that are derived from the existing literature. These are generally concerning the performance during coordination and the stability of new operating points resulting from activating the curative measure. Following \cite{Kolster_2022} among others, the most relevant requirements to the approach presented in this work are:
\begin{enumerate}
    \item \emph{Set point disaggregation:} The requested flexibility for the curative measure needs to be provided by FPU that are distributed within the flexibility providing grid. Activation of a curative measure must result in new set points for these individual units.
    \item \emph{Speed and Reliability:} Curative flexibility must be provided within a preferably short time frame ($\Delta t \leq 2 min$). This includes the disaggregation of new set points to FPU.
    \item \emph{Safe flexibility potentials:} The set points that are being disaggregated to the FPU in the grid must not lead to violations in voltage limits for each bus and operating resource in the respective grid.
\end{enumerate}
The laboratory evaluation of OFO as a coordination algorithm was carried out in regard to the requirements formulated above.

\subsection{Problem Formulation}
\label{subsec:Form}
An optimization problem for the coordination of requested flexibility is formulated and implemented in the OFO controller allowing for an efficient dispatch under the operational constraints of the grid. The concrete optimization problem for the use case was chosen as:
\begin{equation} \label{eq:opti_prob}
\begin{aligned}
    \min              &\quad&\Phi =  \sum_{i \in F} P_{i}^{2} + Q_{i}^{2}  &&     &&     & \\
    \text{s. t.}       &\quad&   V_{min,n}   &\leq V_{meas,n} & \leq V_{max,n} &\quad \forall n \in N\\
                        &\quad&   P_{min,i}   &\leq P_{i} & \leq P_{max,i} &\quad \forall i \in F\\
                        &\quad&   Q_{min,i}   &\leq Q_{i} & \leq Q_{max,i} &\quad \forall i \in F\\
                        &\quad&  P_{PCC} & =P_{set} &
\end{aligned}
\end{equation}
The first three constraints are enforcing the operational limits for the voltages at all buses $N$ and the active and reactive power capabilities of the FPU $F$ in the distribution grid. For the laboratory demonstration of the OFO algorithm in this paper and to showcase its ability to satisfy operational constraints, the feasible voltage band is set to ${V_{\text{min/max}} = V_{\text{N}} \pm 5\%}$. The flexibility demand $P_{\text{set}}$ at the PCC is enforced by implementing it directly into the constraints of the optimization problem. The OFO algorithm is thereby controlling the exchange of active power with the superimposed grid layer by controlling the individual set points for the FPU in the grid. To ensure that the operational limits are met, OFO is additionally controlling reactive power as a degree of freedom in case the bus voltages are exceeding the set limit. The objective function \eqref{eq:opti_prob} for the optimization problem is minimizing the sum of active and reactive power that is feed-in by the FPU in the grid.
 
\subsection{Online Feedback Optimization}
\begin{figure}[b]
    \vspace{-1em}
    \includegraphics[width=0.485\textwidth]{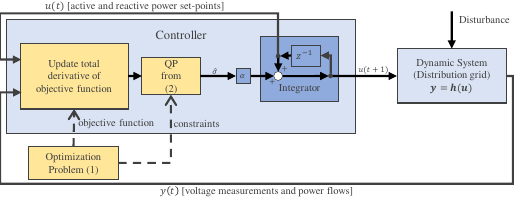}
    \caption{Control loop: OFO controller in a closed loop with distribution grid.}
    \label{fig:ofo_controller}
\end{figure}

This section presents the implemented OFO controller utilized for the dispatch of requested distribution grid flexibility as a curative measure according to the optimization problem described above. The general control loop is shown in \autoref{fig:ofo_controller} and was implemented based on \cite{Härberle_2020}.
The dynamic system that is shown in \autoref{fig:ofo_controller} corresponds for the given use case to the physical distribution grid that is providing the flexibility for the curative measure. This includes all controllable assets and operational resources, as well as external disturbances such as uncertainties in voltage at the PCC or varying generation and consumption. The physical system is in a closed loop by incorporating online measurements from the grid in each iteration of the algorithm, with $u \in \mathbb{R}^{p}$ being the set points for active and reactive power, $y \in \mathbb{R}^{n}$ the measurements for voltage and power flow from the grid and $\alpha \in \mathbb{R}^{>0}$ denoting the fixed step-size parameter for the controller. For a given objective function $\Phi(u,y): \mathbb{R}^{p} \times \mathbb{R}^{n} \rightarrow \mathbb{R}$, OFO is aiming to find the optimal solution considering equality and inequality constraints. A single step of OFO includes:
\begin{enumerate}
    \item Acquiring the feedback values for the next iteration of the controller through measurements from the grid.
    \item Calculating the gradient of the objective function chosen for the optimization problem \eqref{eq:opti_prob} based on momentary values for $u$ and $y$.
    \item Calculating $\hat{\sigma}(u,y)$ by projecting the gradient of the objective function onto the set of feasible set points by solving an internal quadratic program (QP) \eqref{eq:ofostepcalculation} to find the best constrained solution for the iteration.
    \item Calculating the next set point vector with ${u(t+1) = u(t) + \alpha\hat{\sigma}(u,y)}$.
    \item Sending set points for active and reactive power to the FPU.
\end{enumerate}
With $\hat{\sigma}(u,y)$ being the solution of the internal QP that is used to project the gradient of the objective function  ${H}(u)^{T}\nabla \Phi(u,y)$ onto the set of constrained set points.
\begin{equation} \label{eq:ofostepcalculation}
	\begin{aligned}
		\hat{\sigma}(u,y) \coloneqq \arg \min_{w \in \mathbb{R}^{p}} \quad & \| w + H(u)^{T}\nabla \Phi(u,y)\|^{2}\\
		\textrm{s. t.} \quad & \begin{bmatrix}P_{min,i} \\Q_{min,i}\end{bmatrix}\leq \begin{bmatrix}P_i \\Q_i\end{bmatrix} + \alpha w \leq \begin{bmatrix}P_{max,i} \\Q_{max,i}\end{bmatrix}\\
		\quad & P_{set} = P_{PCC}(w)\\
		\quad & V_{min} \leq V_{meas} + \alpha \nabla h(u) w \leq V_{max} \\\\
		\textrm{with} \quad &H(u)^{T} \coloneqq [ \mathbb{I}_{p} \hspace{2mm} \nabla h(u)^{T}]\\
        \textrm{and}  \quad & w \coloneqq \begin{bmatrix}
			                 \Delta P\\ \Delta Q
		                      \end{bmatrix}
\end{aligned}
\end{equation} 
The sensitivity matrix $\nabla h(u)$ is determined a priori and represents a steady-state input-output sensitivity map of the physical system as the only explicit model information that is needed for the OFO algorithm. It describes how a change in active or reactive power changes the bus voltages and is therefore similar to power transfer distribution factors (PTDF). It only has to be calculated once for a fixed set point vector $u$, because the algorithm in itself is robust against model mismatch within the sensitivity matrix. The algorithm runs continuously over multiple iterations which results in the dynamic system converging to a state that is optimal in the sense of the optimization problem \eqref{eq:opti_prob}. By incorporating online measurements as feedback in the control loop instead of using explicit power system models, the OFO algorithm holds several advantages over conventional coordination algorithms, i.e. \cite{Hau_2021}:
\begin{enumerate}
    \item Very little explicit model information are needed
    \item Robustness against model mismatch and time-varying disturbances
    \item Computational efficiency
\end{enumerate}
OFO is therefore, in its presented implementation, implicitly fulfilling the requirements for curative flexibility provision (see subsection \ref{subsec:Requirements}) and suitable for the general type of dispatch problem to be solved for the presented use-case. To further underline its applicability we deploy the algorithm in an experimental laboratory set up to evaluate it for possible limitations that might arise in grid operation. These include external disturbances, as well as technical limitations like delay and sampling time of measurements and communication with FPU in the hardware setup.

\section{Laboratory Setup}
\label{sec:Labor}
\begin{figure}[b]
    \includegraphics[width=0.5\textwidth]{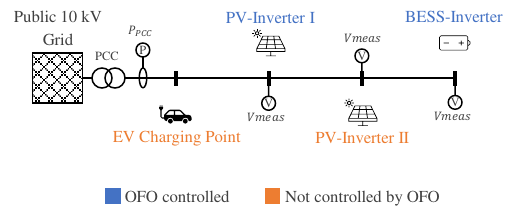}
    \caption{Full laboratory setup used for experimental validation.}
    \label{fig:lab_setup}
\end{figure}

To showcase the method for flexibility coordination and to validate its capabilities in terms of the requirements of curative system operation, it was implemented in an exemplary low voltage grid with a single PCC to the public medium voltage grid. The setup includes grid following inverters receiving set points for active and reactive power by the OFO algorithm, as well as an independent inverter utilizing a $Q(V)$ control and an electric vehicle (EV) charging point. The laboratory setup is shown in \autoref{fig:lab_setup}. The controller acts as a central operating logic, iteratively receiving measurements from the grid and sending set points to the controllable inverters with a sampling time of $t_{\text{sample}} = 5s$. Coordination of inverters and handling of data during the iterative process is done using a co-simulation approach running parallel to the physical laboratory \cite{Schmidtke_2022, Hacker_2021}.
The described setup is inherently containing several possible external disturbances that are influencing the OFO controller during flexibility coordination by additionally influencing the operating point at each bus in the system. These include:
\begin{enumerate}
    \item Varying voltage levels at the PCC
    \item $Q(V)$ control in legacy inverters
    \item Varying power consumption due to EV charging
\end{enumerate}
During the experiments that were carried out for this work the OFO algorithm was tested for robustness against any of the disturbances mentioned above to investigate if the chosen approach is feasible for flexibility provision in curative system operation. The results of these experiments are described in detail in the next section.

\section{Exemplary Results}
\label{sec:Results}
\subsection{Time-frame of Flexibility Provision}
\begin{figure}[b]
    \vspace{-1em}
    \input{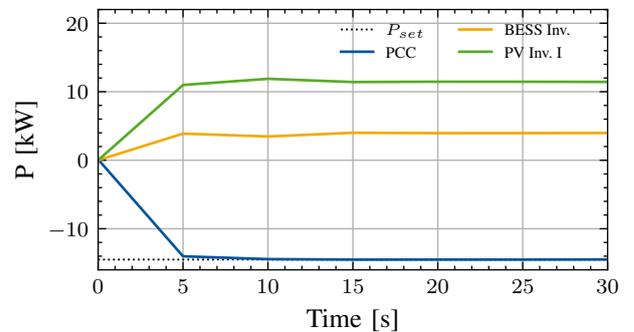}
    \caption{Coordination of $14.5 \text{kW}$ flexible active power by OFO algorithm.}
    \label{fig:res_base_coord}
\end{figure}

To address research question i. (see \ref{subsec:RQ}) and the requirement for \emph{speed and reliability} in coordination (see \ref{subsec:Requirements}) a flexibility demand of $P_{\text{set}} = -14.5 \text{kW}$ is disaggregated to the BESS inverter and PV inverter I. The time $\Delta t$ is measured that passes before the requested set point is fully reached by the controller. No external disturbances aside from possible fluctuation in the voltage level of the public grid are considered. The results of the coordination are shown in \autoref{fig:res_base_coord}. It is visible, that the requested amount of active power at the PCC (blue) is reached within three iterations of the OFO algorithm after $\Delta t=10s$ for the chosen sampling time and remains without tracking error afterwards. The coordination of flexibility via OFO is therefore a viable option for an exemplary distribution grid in a time frame that is relevant to curative system operation.

Furthermore, it is visible that the controllable inverters are actuated differently by the OFO controller. The PV inverter is injecting $11.5 \text{kW}$ while the BESS inverter is only injecting $4 \text{kW}$. This behavior during coordination is a direct result of the chosen objective function in the optimization problem \eqref{eq:opti_prob} minimizing active and reactive power feed-in during flexibility provision. Higher active power feed-in by the BESS would result in an overall higher need for reactive power to keep bus voltages in the permissible range. The OFO algorithm is consequently choosing assets that are positioned within a shorter electrical distance to the PCC without this information being available in the form of a concrete model of the distribution grid. Therefore, the second requirement for curative flexibility \emph{set point disaggregation} as formulated in \autoref{subsec:Requirements} is fulfilled through the presented choice of objective function.

\subsection{Robustness against Disturbances}
\begin{figure}[t]
    \input{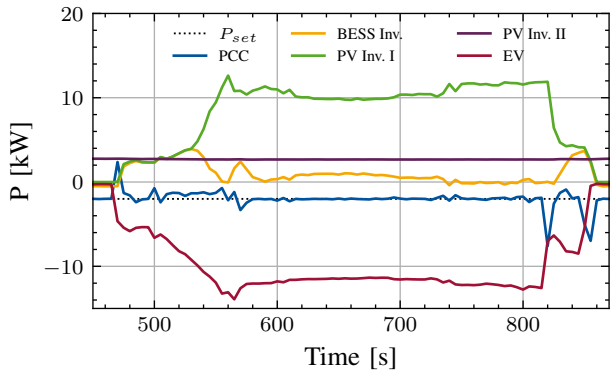}
    \caption{OFO controller in operation under external disturbances.}
    \label{fig:ofo_op}
    \vspace{-1em}
\end{figure}

\begin{figure}[t]
    \input{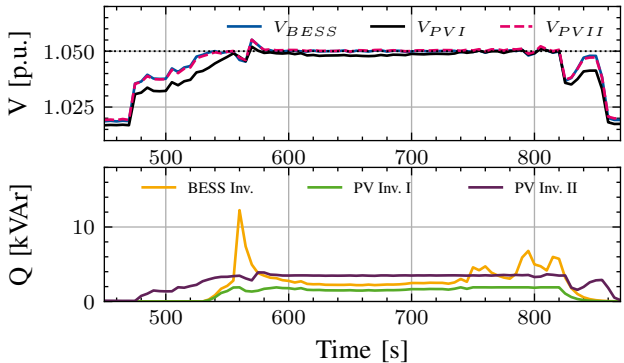}
    \caption{Voltage control by OFO to ensure constraints.}
    \label{fig:ofo_op_q}
    \vspace{-1.5em}
\end{figure}

Research question ii. (see \ref{subsec:RQ}) is addressed by testing the implemented OFO controller under circumstances that are designed to emulate online grid operation. The algorithm is controlling the flow of active power $P_{\text{set}} = -2 \text{kW}$, at the PCC of the distribution grid as shown in \autoref{fig:lab_setup}. The robustness of the algorithm against external disturbances by actors in the grid that are unknown to the controller is investigated. In the given scenario these include an independent PV inverter and an EV that is starting its charging process with up to ${P_{charge} = -14 \text{kW}}$ at $t=470s$. It is visible in \autoref{fig:ofo_op} that the measured active power at the PCC is shortly influenced by the added consumption through the charging process of the EV, but brought back to the requested set point within a few seconds. The additional active power is again predominantly provided by the PV inverter. The bus voltages are kept in the permissible range as shown in \autoref{fig:ofo_op_q}. The OFO-controlled inverters, as well as the independent PV inverter, are reacting to the increased voltage with a feed-in of reactive power. Visibly, the independent inverter provides reactive power according to its internal $Q(V)$ controller. The OFO-controlled assets are following shortly after with higher gradients, especially in the BESS inverter. Sporadically the bus voltages are rising slightly above the permitted range but are brought back by the OFO controller within a few iterations. This is mainly due to the chosen sampling rate for the measurements that are being processed by OFO and the delay times present in the hardware setup and therefore is not an inherent limitation of the OFO algorithm. The implemented controller is exhibiting a high level of robustness against external disturbances that can occur during flexibility provision.

\section{Conclusion}
\label{sec:Conclusion}
In this paper we introduced a method to coordinate distribution grid flexibility for curative system operation by using OFO. The controller was implemented in a distribution grid laboratory and evaluated in terms of its performance for the specific use case. It was shown, that OFO is holding significant advantages over conventional approaches and is a viable option to coordinate flexibility for a curative measure, due to its speed and reliability under external disturbances. This is underlined by the low need for information about the grid and the robustness against model mismatch that sets this method further apart from conventional OPF calculations. In future work, possible interference between multiple OFO controllers during coordination will be investigated. This situation can occur when different grid operators use a cascaded hierarchy of OFO controllers to provide flexibility over several layers of distribution grids in case of an overloaded transmission line.

\section*{Acknowledgment}
\begin{wrapfigure}{r}{0.12\textwidth}
  \vspace{-\baselineskip}
  \vspace{-\baselineskip}
  \includegraphics[width=0.12\textwidth]{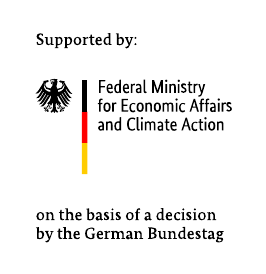}
  \vspace{-\baselineskip}
    \vspace{-\baselineskip}
        \vspace{-\baselineskip}
\end{wrapfigure}
This project received funding from the German Federal Ministry for Economic Affairs and Climate Action under the agreement no. 03EI4046E (PROGRESS).

\end{document}